\documentclass[osajnl,twocolumn,showpacs,superscriptaddress,10pt]{revtex4-1} %% use 11pt for Applied Optics
\usepackage{amsmath,amssymb,graphicx}

\renewcommand{\eqref}[1]{Eq.~(\ref{#1})}

\def\figref{Fig.~\ref}
\def\figureref{Figure \ref}

\begin{document}

\title{Larger-area single-mode photonic crystal surface-emitting lasers\\ enabled by an accidental Dirac point}

\author{Song-Liang Chua}\email{Corresponding author: csonglia@dso.org.sg}
\affiliation{Research Laboratory of Electronics, Massachusetts Institute of Technology, Cambridge, Massachusetts 02139, USA}
\affiliation{DSO National Laboratories, 20 Science Park Drive, Singapore 118230, Singapore}

\author{Ling Lu}\email{Corresponding author: linglu@mit.edu}
\affiliation{Research Laboratory of Electronics, Massachusetts Institute of Technology, Cambridge, Massachusetts 02139, USA}

\author{Jorge Bravo-Abad}
\affiliation{Departamento de F\'\i sica Te\'orica de la Materia Condensada and Condensed Matter Physics Center (IFIMAC), Universidad Aut\'onoma de Madrid, 28049 Madrid, Spain}

\author{John D.~Joannopoulos}
\author{Marin Solja\v{c}i\'{c}}
\affiliation{Research Laboratory of Electronics, Massachusetts Institute of Technology, Cambridge, Massachusetts 02139, USA}

\begin{abstract}By altering the lattice geometry of the photonic crystal surface-emitting lasers (PCSELs), we tune the regular lasing band edges of quadratic dispersions to form a single accidental Dirac point of linear dispersion at the Brillouin zone center.
This not only increases the mode spacing by orders of magnitude, but also eliminates the distributed in-plane feedback to enable single-mode PCSELs of substantially larger area thus substantially higher output power.
The advantages of using accidental Dirac cones are systematically evaluated through two-dimensional in-plane calculations and confirmed by three-dimensional simulations of photonic crystal slab devices.
\end{abstract}

%\ocis{(140.3490) Lasers, distributed-feedback; (250.7270) Vertical emitting lasers; (230.5298) Photonic crystals.}% REPLACE WITH CORRECT OCIS CODES FOR YOUR ARTICLE % NOTE: \ocis{} IS ALIASED TO \pacs{} BUT MUST FORMAT THE TERMS CORRECTLY FOR EACH JOURNAL

\maketitle %% required

%%%%%%%%%%%%%%%%%%%%%%%%%%  body  %%%%%%%%%%%%%%%%%%%%%%%%%%
Higher power single-mode on-chip lasers with good beam qualities are of interest for many applications.
While the edge emitting sources (distributed feedback lasers) suffer from catastrophic optical damage at their facets, surface-emitting sources (vertical cavity surface-emitting lasers) are usually limited by their small cavity sizes.
In both examples, the single lasing mode is selected by means of one-dimensional feedback structures.
Utilizing two-dimensional distributed feedback, surface emitters have achieved broad-area single-mode operations \cite{Imada02,Sun08,Chassagneux09}.
In particular, PCSELs have not only achieved the highest surface-emitting single-mode power \cite{Kunishi06} but also the ability to control the shapes \cite{Miyai06}, polarizations \cite{Noda01} and directions \cite{Kurosaka10} of the laser beams.
PCSELs are essentially the two-dimensional (2D) versions of the second-order distributed feedback lasers \cite{Henry85}, where the higher quality factor lasing mode is selected through the symmetry mismatch to the free-space modes \cite{Ochiai01,Fan02,FanoPaper}.
However, the lasing areas of PCSELs are limited by two fundamental constraints.
Firstly, the mode spacing decreases as the cavity area increases, which promotes multi-mode lasing.
Secondly, the distributed in-plane feedback localizes the lasing fields to individual coherent sections, which promotes multi-area lasing.

In this work, we tune the regular lasing band edges of quadratic dispersions to form accidental Dirac cones \cite{Huang11,Sakoda12,SongIEEE12} of linear dispersions.
This not only increases the mode spacing by orders of magnitude but also eliminates the distributed in-plane feedback, turning the periodic index-modulated cavities into equivalent Fabry-Perot-like cavities where the modes, however, have different out-of-plane coupling losses.
Both of these advantages lead to single-mode PCSELs of significantly larger areas and thus higher output powers.
In the following, we systematically evaluate the advantages of using accidental Dirac cones through 2D calculations and provide consistent results of 3D slab devices.
The influences of the photonic crystal (PhC) structures on the PCSELs' lasing characteristics can be accounted for in terms of the modal properties and dispersions of the corresponding passive systems \cite{FanoPaper,Liang11}.
Thus, in this work, we focus on analyzing the passive characteristics of the PhCs.

A Dirac cone is a special dispersion relation in the band structure where the dispersion is linear and the density of states (DOS) vanishes at the Dirac point.
It is well-known that pairs of Dirac cones can exist in PhCs \cite{Plihal91,Chong08}.
However, a single Dirac cone at the Brillouin zone center $(\Gamma)$ is ideal for both vertical emission and single-mode lasing in a PCSEL.
This single Dirac cone can form when the PhC is geometrically tuned so that a singly-degenerate band is accidentally degenerate with a pair of doubly-degenerate bands at $\Gamma$ \cite{Huang11,Sakoda12}.
When this happens, two of the three bands form an isotropic \cite{Urzhumov05} Dirac cone and the other one is flat.
Such accidental Dirac points can universally exist in PhCs of square or triangular lattices consisting of either dielectric-rod or air-hole arrays, with either high or low index contrasts.

\begin{figure}[t]
\centering\includegraphics[width=0.49\textwidth]{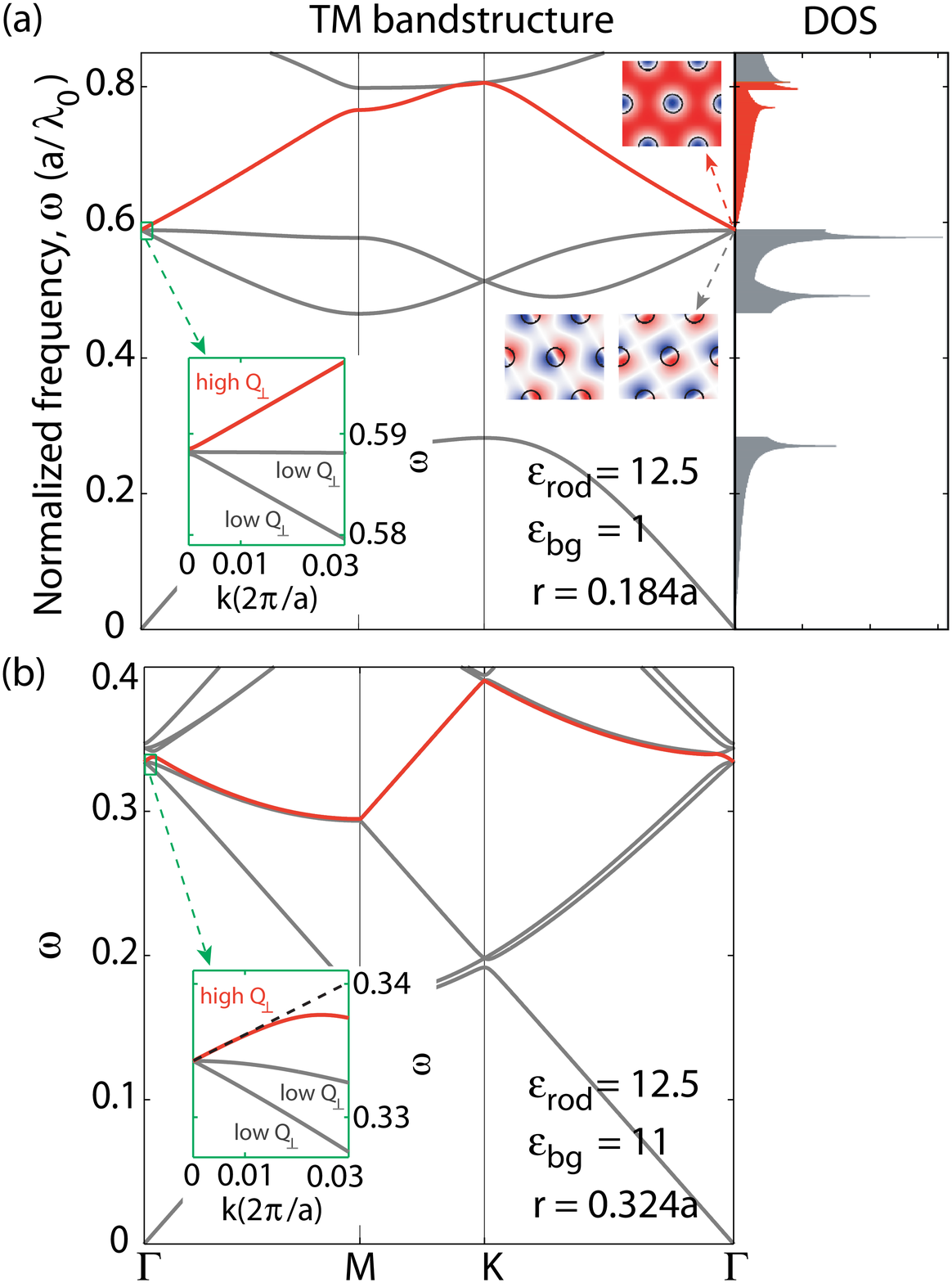}
\caption{\small{TM photonic band structure of a 2D triangular array of dielectric rods $(\epsilon_{\text{rod}}=12.5)$ in two different background materials $(\epsilon_{\text{bg}}=1 \text{ and } 11)$. In both cases, rod radius $r$ is tuned so that the doubly-degenerate modes (with $E_1$ symmetry in the $C_{6_v}$ point group) are accidentally degenerate with a singly-degenerate $A_1$ mode (red band) at $\Gamma$. (a) The band structure and DOS of the high dielectric contrast PhC. The modal profiles of the three accidentally degenerate modes at $\Gamma$ are depicted with electric field pointing into the page and having positive (negative) values in red (blue). (b) The band structure of the low dielectric contrast PhC.}} \label{fig:SONG_1}
\end{figure}

In \figref{fig:SONG_1}, we obtained the single accidental Dirac cone in a 2D triangular array of dielectric rods by tuning the rod radius $r$, where the lattice constant is denoted by $a$.
The calculations were performed with a unit-cell using the MIT Photonic-Bands package \cite{Johnson01}.
To show the generality of our approach, we have considered PhCs with both a high and low dielectric contrast; the dielectric contrast is high ($12.5:1$) in \figref{fig:SONG_1}(a) and low ($12.5:11$) in \figref{fig:SONG_1}(b).
The resulting linear dispersion of the singly-degenerate mode (red band) has DOS that vanishes linearly with frequency at the accidental Dirac point, as shown in the right plot of \figref{fig:SONG_1}(a).
This corresponds to large mode spacings near the band edge and a high spontaneous emission coupling factor \cite{Jorge12}.
Modal profiles depicting the point group symmetry characteristic to each of the three accidentally degenerate modes at $\Gamma$ are also illustrated in \figref{fig:SONG_1}(a).
The simulations shown in \figref{fig:SONG_1} are 2D.
However, when a 2D PhC slab is implemented using the simulations as a guidance, the modes at $\Gamma$ (being above the light-line) will be able to couple to out-of-plane radiation.
Only the symmetry of the singly-degenerate mode is mismatched with the free space modes \cite{Ochiai01} and so, has lower out-of-plane radiation losses (higher $Q_{\perp}$) than the other two modes.
Thus, the band edge modes of this singly-degenerate band are the only lasing candidates within the spectral range shown in the insets of \figref{fig:SONG_1}(a) and (b).
From now on, we only consider the modes of this singly-degenerate band.

\begin{figure}[t]
\centering\includegraphics[width=0.485\textwidth]{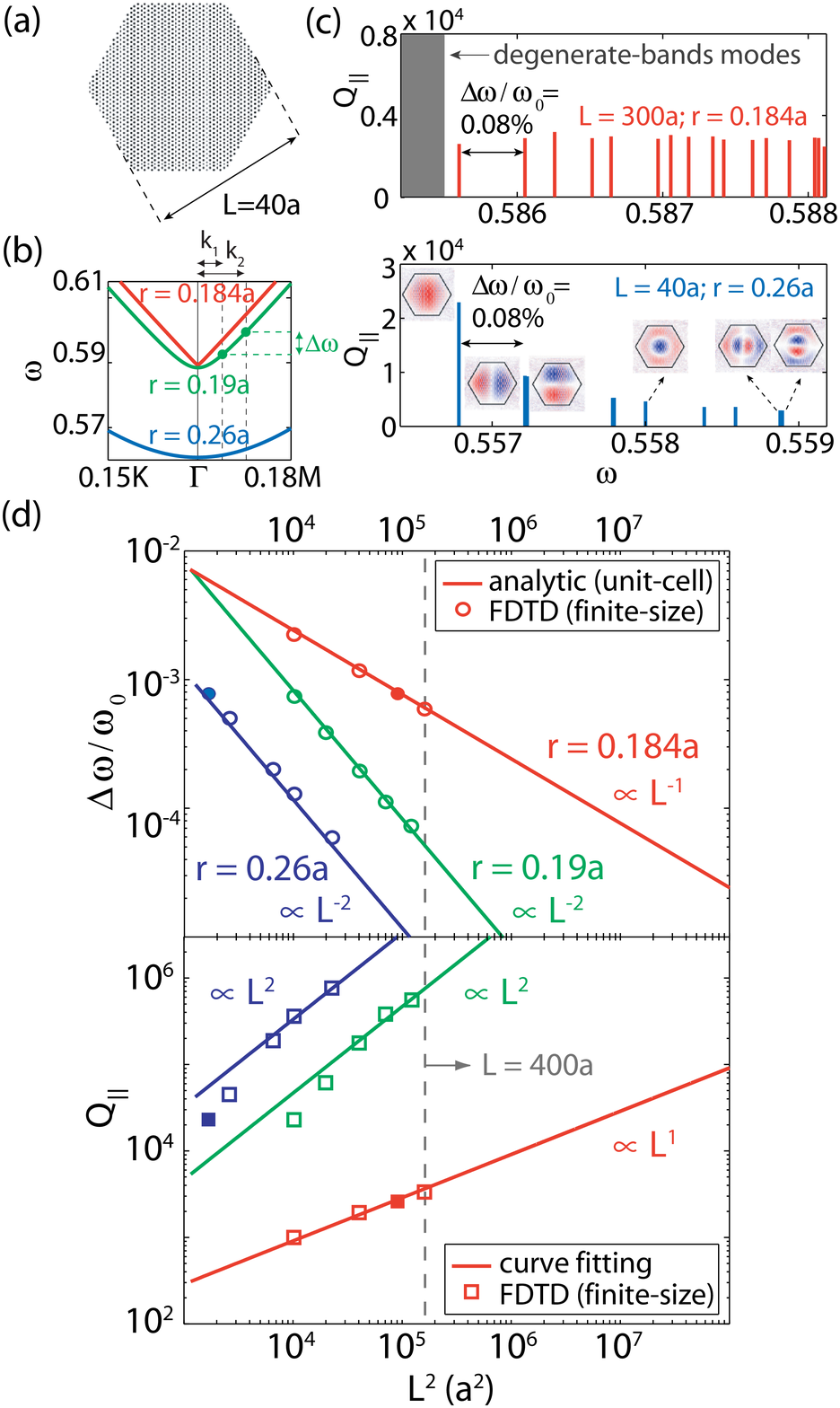}
\caption{\small{(a) A finite-sized 2D PhC cavity consisting of a triangular array of dielectric rods $(\epsilon_{\text{rod}}=12.5)$ embedded in air with dimension $L=40a$. (b) The high $Q_{\perp}$ band dispersions of three PhCs with different rod radii. At $\Gamma$, the red band has a linear dispersion while the green and blue bands have quadratic dispersions. (c) Band edge modes of two finite-sized PhC cavities of $L=300a$ with linear dispersion (top plot) and of $L=40a$ with quadratic dispersion (lower plot). Insets show the mode profiles with electric field pointing into the page and having positive (negative) values in red (blue). (d) The upper plot illustrates the band edge mode spacing ($\Delta\omega/\omega_0$ where $\omega_0$ is the band edge frequency) as a function of the cavity area. The lower plot illustrates $Q_{\parallel}$ of the band edge mode as a function of the cavity area. The filled circles and squares correspond to the results shown in (c).}} \label{fig:SONG_2}
\end{figure}

To quantify the benefits of having linear dispersions at the lasing band edges, we first compare the band edge modes of two finite-sized 2D PhCs.
One of them has linear dispersion at the accidental Dirac point, while the other has quadratic dispersion.
The PhC analyzed is a triangular array of dielectric rods ($\epsilon_{\text{rod}}$=12.5) in air that is truncated with a hexagonal-shaped boundary of dimension $L$ shown in \figref{fig:SONG_2}(a).
We altered the band dispersions of the PhC by varying the rod radius $r$.
The dispersion relations for three values of $r$ are illustrated in \figref{fig:SONG_2}(b).
A linear dispersion is formed when $r=0.184a$ while quadratic dispersions are formed at the remaining two radii.
In \figref{fig:SONG_2}(c), the quality factors $Q_{\parallel}$ of the band edge modes are plotted as a function of the frequency for a cavity of $L=300a$ with linear dispersion (top plot) and a cavity of $L=40a$ with quadratic dispersion (lower plot).
These results were calculated using the finite-difference time-domain (FDTD) method \cite{Oskooi10} with perfectly matched layer boundary regions. 
Remarkably, the $300a$ PhC cavity at the accidental degeneracy is found to have the same mode spacing (between the first two band edge modes) as that of the $40a$ cavity even though the former is $50$ times larger in area.
Moreover, $Q_{\parallel}$ of the first band edge mode in the $300a$ cavity (linear dispersion) is an order of magnitude smaller than that in the $40a$ cavity (quadratic dispersion); this is because the group velocity ($v_g=d\omega/dk$) of the first band edge mode is larger in the linear dispersion case.
We note that the low $Q_{\parallel}$ implies a weak in-plane feedback and low localization effects.

Another distinction between the cavities with linear and quadratic dispersions in \figref{fig:SONG_2}(c) is that the modes of the top plot (linear dispersion) all have similar $Q_{\parallel}$ values (since $v_g$ is constant) while the modes of the lower plot (quadratic dispersion) have $Q_{\parallel}$ values that are decreasing with frequency (since $v_g$ is increasing).
Nonetheless, in both cases, the first band edge modes having the highest $Q_{\perp}$ values will lase with the lowest thresholds, because $Q_{\parallel}$ values increase with the cavity sizes and will not be the dominant loss mechanism in large cavities.

In the upper plot of \figref{fig:SONG_2}(d), we compare the mode spacing as a function of the PhC area for linear and quadratic dispersions at the band edge.
The data points in the figure are results obtained from FDTD calculations of the finite-sized PhC cavities.
The solid lines are plotted using analytic expressions with dispersion curvatures extracted from the band structures in \figref{fig:SONG_2}(b).
The two approaches are consistent with each other.
To get the analytic behavior, we assume that the first and the second band edge modes have the same frequencies as the modes of wavevectors $k_1=\pi/L$ and $k_2=2\pi/L$ in the infinite system, because they have similar mode profiles and boundary conditions, as shown in the lower plot of \figref{fig:SONG_2}(c).
The band edge mode spacing is then the frequency difference between these two $k$-points illustrated in \figref{fig:SONG_2}(b).
Using this approach, the mode spacing ($\Delta\omega$) of a linear dispersion is found to be inversely proportional to $L$ ($\Delta\omega=\pi\beta/L$) while the mode spacing of a quadratic dispersion is inversely proportional to $L^2$ ($\Delta\omega=3\pi^2\alpha/L^2$).
$\beta$ is the linear slope and $\alpha$ is the quadratic curvature of the dispersions near the band edge.
These semi-analytical expressions are verified by the mode spacings from FDTD simulations shown in the upper plot of \figref{fig:SONG_2}(d).

The results clearly indicate that by tuning to a linear dispersion, the mode spacing can be made much larger than that in a typical PCSEL with quadratic dispersion.
For instance, at $L=400a$ [dashed vertical line in \figref{fig:SONG_2}(d)], the mode spacing at the accidental point (red line) is at least $60$ times larger compared to the PhC detuned from it (blue line).
We note that the difference in the mode spacings between the linear and the quadratic case becomes arbitrarily large as the area increases.
Equivalently, for the same mode spacing, the PhCs with a linear dispersion can be made much larger in area than those with quadratic dispersions.
In \figref{fig:SONG_2}(d), the cavity size is increased by more than two orders of magnitude when $r$ is tuned from $0.26a$ to $0.184a$ while maintaining the same mode spacing ($\Delta\omega/\omega_0=1\times10^{-4}$).

In the lower plot of \figref{fig:SONG_2}(d), we compare the in-plane feedback as a function of the PhC cavity area for linear and quadratic dispersions at the band edge.
We quantify the in-plane feedback strength by the in-plane quality factor $Q_{\parallel}=\omega_{0}\tau_{\parallel}$, where $\tau_{\parallel}$ $(\propto L/v_g)$ is the photon lifetime in the PhC cavity.
$v_g$ is a constant when the dispersion is linear, and is proportional to $k$ when the dispersion is quadratic.
The above analysis on mode spacing finds that $k$ scales as $1/L$.
Hence, $Q_{\parallel}$ should scale with $L$ when the dispersion is linear and $L^2$ when the dispersion is quadratic.
In \figref{fig:SONG_2}(d), $Q_{\parallel}$ calculated from finite-sized cavities agree well with the above trends, except for small structures whose modes are of $k$ values too far away from $\Gamma$ to follow the quadratic functions.
Physically, the linear increase of $Q_{\parallel}$ with $L$ implies that the distributed in-plane feedback in a typical PCSEL is completely eliminated at the accidental degeneracy.
In other words, the PCSEL behaves like a 2D Fabry-Perot cavity where feedback only comes from its end mirrors.
However, unlike typical Fabry-Perot cavities where all the modes have the same $Q$ values, the proposed PCSELs can still select the first band edge mode to lase due to its highest $Q_{\perp}$ value.
In real devices, fabrication imperfections can cause field localization effects prohibiting coherent lasing over a larger area, but those effects will be much reduced when the band edge has linear dispersion.

\begin{figure}[t]
\centering\includegraphics[width=0.48\textwidth]{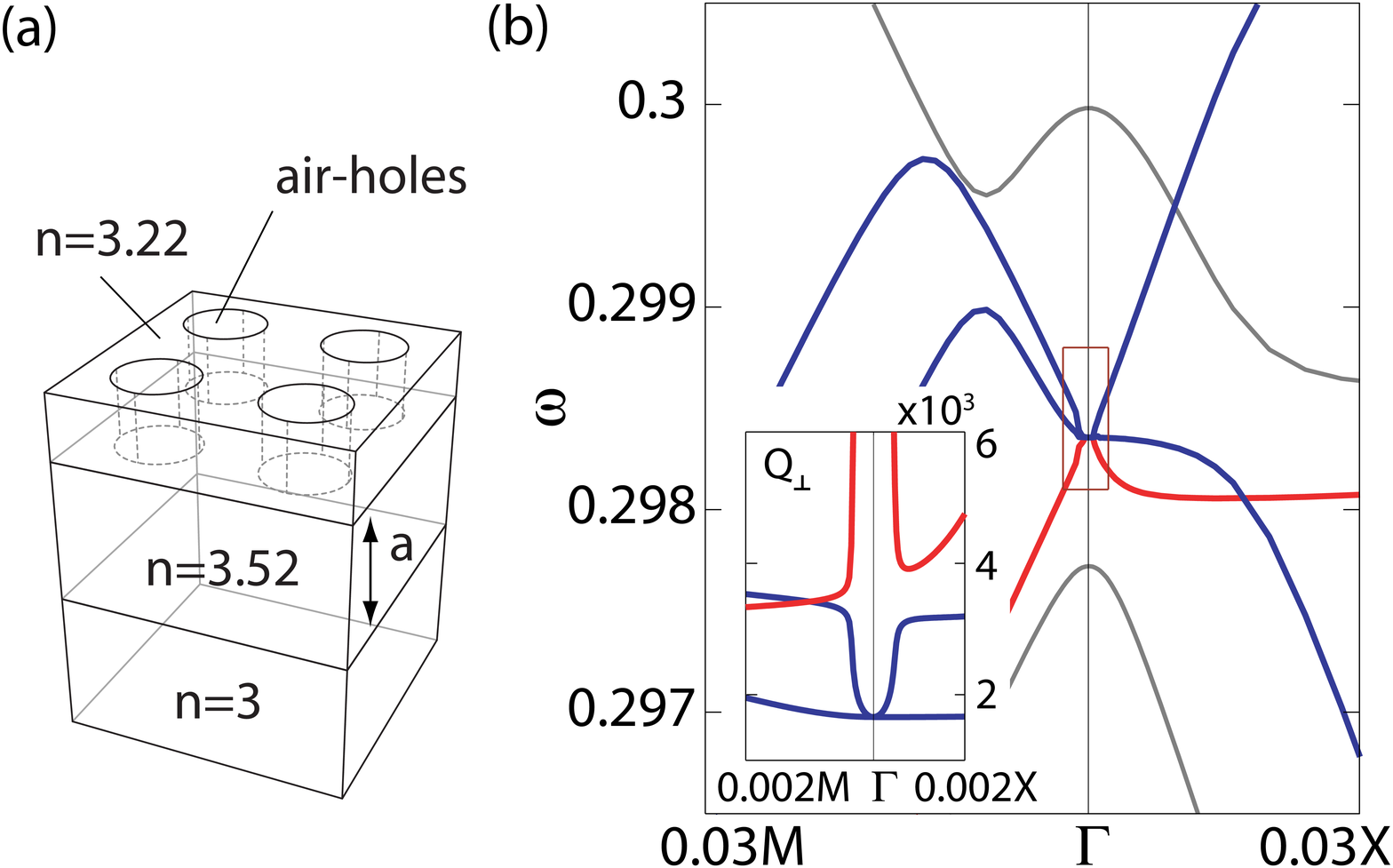}
\caption{\small{(a) $2\times2$ unit-cells of a GaAs-based 2D PhC slab on AlAs $(n=3)$ substrate.
A square array of air-holes with radius $0.285a$ is patterned in the top $0.47a$-thick GaInP $(n=3.22)$ layer.
Sandwiched between GaInP and the substrate is a homogeneous $1a$-thick GaAs $(n=3.52)$ layer.
$a$ is the lattice constant.
(b) Band structure (TE-like modes near $\Gamma$) of the considered PhC slab. The hole radius is tuned so that the pair of doubly-degenerate modes (blue bands) is accidentally degenerate with a singly-degenerate mode (red) at $\Gamma$.
Bottom inset shows the corresponding Q-plot of the three bands in a small region near $\Gamma$, where $Q_{\perp}$ diverges only for the red band.
}} \label{fig:SONG_4}
\end{figure}

The above analysis thus far has considered 2D PhCs without out-of-plane radiation losses.
We now consider a 2D PhC GaAs-based slab on AlAs substrate with open boundaries in the vertical directions [see sketch in \figref{fig:SONG_4}(a)].
\figureref{fig:SONG_4}(b) illustrates its band structure near $\Gamma$ when the air-hole radius is tuned to achieve an accidental degeneracy.
The inset shows the corresponding $Q_{\perp}$ values of the three bands enclosed by the red box, in a small region near $\Gamma$.
These calculations are performed with the finite element method using the commercially available software COMSOL.
Similar to 2D PhCs, linear dispersions of two bands form in the PhC slabs when three band edges are tuned to be degenerate.
The linear dispersions imply that band edge modes with both large mode spacings and low in-plane feedback can be realized.
Single-mode PCSEL operation is possible because $Q_{\perp}$ diverges only for one (red) band and its $Q_{\perp}$ value drops rapidly as $k$ deviates from $\Gamma$, as shown in the inset of \figref{fig:SONG_4}(b).
Interestingly, if one examines very close to $\Gamma$ $\left[\text{i.e.~}|k|<0.0005(2\pi/a)\right]$, a very different dispersion relation exists. Such occurrence, however, does not affect the conclusion of this work and will be the subject of a different paper.

In conclusion, we have demonstrated that, compared to typical PCSELs with quadratic band edge dispersions, the formation of accidental Dirac cones of linear dispersions at $\Gamma$ not only increases the mode spacing by orders of magnitude but also eliminates the distributed feedback in-plane. 
This overcomes two of the fundamental limitations to attaining larger-area and higher-power single-mode PCSELs.

We acknowledge helpful discussions with Prof.~Steven Johnson and Prof.~A.~Douglas Stone.
S.L.C.~was supported in part by the MRSEC Program of the NSF under Award No.~DMR-0819762. This work was also supported in part by the Army Research Office through the ISN under Contract W911NF-13-D-0001. L.L.~and M.S.~were supported in part by the MIT S3TEC Energy Research Frontier Center of the Department of Energy under Grant DE-SC0001299. S.L.C.~acknowledges the financial support from the DSO National Laboratories, Singapore. J.B.A.~acknowledges financial support by the Ramon-y-Cajal program, Grant No. RyC-2009-05489.

\end{document}